\documentclass[twocolumn,amsmath,amssymb,pre]{revtex4}

\usepackage{graphicx}
\usepackage{dcolumn}
\usepackage{bm}

\begin{document}



\title{Critical exponents from cluster coefficients}

\author{Z. Rotman}
\email{rotmanz@gmail.com}
\author{E. Eisenberg}
\email{elieis@post.tau.ac.il}

\affiliation{Raymond and Beverly Sackler School of Physics and Astronomy,
Tel Aviv University, Tel Aviv 69978, Israel}



\begin{abstract}
For a large class of repulsive interaction models, the Mayer cluster
integrals can be transformed into a tridiagonal real symmetric
matrix $R_{mn}$, whose elements converge to two constants. This
allows for an effective extrapolation of the equation of state for
these models. Due to a nearby (nonphysical) singularity on the
negative real $z$ axis, standard methods (e.g. Pad\`{e} approximants
based on the cluster integrals expansion) fail to capture the
behavior of these models near the ordering transition, and, in
particular, do not detect the critical point. A recent work
(Eisenberg and Baram, PNAS {\bf 104}, 5755 (2007)) has shown that
the critical exponents $\sigma$ and $\sigma'$, characterizing the
singularity of the density as a function of the activity, can be
exactly calculated if the decay of the $R$ matrix elements to their
asymptotic constant follows a $1/n^2$ law. Here we employ
renormalization arguments to extend this result and analyze cases
for which the asymptotic approach of the $R$ matrix elements towards
their limiting value is of a more general form. The relevant
asymptotic correction terms (in RG sense) are identified and we then
provide a corrected exact formula for the critical exponents. We
identify the limits of usage of the formula, and demonstrate one
physical model which is beyond its range of validity. The new
formula is validated numerically and then applied to analyze a
number of concrete physical models.
\end{abstract}


\maketitle

\section{INTRODUCTION}


It is often said that the mechanism underlying phase transitions is
the decrease of internal energy in the ordered phase. However, it
has been shown long ago that melting is dominated by the strong
short ranged repulsive forces, and the related solid-fluid
transitions are entropy-driven. Accordingly, purely repulsive models
have been often used to study the fluid equation of state towards
the structural ordering transition. The most striking demonstration
of these observations is given by the family of hard-core models,
which have long played a central role in this field. In these
models, particles interact exclusively through an extended hard
core, and there is no temperature scale associated with the
potential (interaction energy is either infinite inside the
exclusion region or zero outside). Thus, temperature and energy play
no role, and the dynamics is completely determined by entropy
considerations. Yet, these models exhibit various types of ordering
transitions. They include, for example, the famous isotropic-nematic
transition in a three dimensional system of thin hard rods
\cite{onsager,zwanzig}, as well as the extensively studied hard
spheres models \cite{wood,alder,alderdisks,hoover,michels}, undergoing a
first order fluid-solid transition for $d\geq 3$ and, presumably,
a second order transition from a fluid to the hexatic phase
\cite{hexatic1,hexatic2}. These models are purely entropy-driven, yet they
capture the essential molecular mechanism that drives freezing
transitions.

A complete description of the fluid phase is provided by the Mayer
cluster series in terms of the activity, $z=\exp(\beta\mu)$, where
$\mu$ is the chemical potential. For purely repulsive potentials,
the radius of convergence of the cluster series is known to be
determined by a singularity on the negative real axis, $z=-z_0$,
typically very close to the origin \cite{Groeneveld}.
Near this point, the singular part of the density is
characterized by the critical exponent $\sigma$:
$$\rho_{\rm sing}(z) \simeq (z+z_0)^\sigma.$$
As a result of this singularity,
the radius of convergence of the Mayer series includes only the
extremely low density regime, and the fluid-solid transition is way
beyond it. It is therefore desirable to find a way to extend the
information contained in the cluster integrals series to provide
information about the behavior of the system close to the ordering
transition region. In particular, one is interested in the critical
exponent $\sigma'$ characterizing the density near the physical
termination point of the fluid $z_t$:
$$\rho_{\rm sing}(z) \simeq (z-z_t)^{\sigma'}.$$

It has been shown that this goal may be achieved by transforming the
cluster integral series into a tridiagonal symmetric matrix form
\cite{baramr}. The matrix elements $R_{nm}$ adopt a clear asymptotic
form, and converge extremely fast to two different constants: $A$
(off-diagonal) and $B$ (diagonal). This fact can then be utilized to
obtain good approximants for the fluid density far outside of the
convergence circle of the power series \cite{baramf,baramr2,eb06}.
Like Pad\`{e} methods, these approximants are consistent with the
known elements to all available orders. However, the $R$ matrix
scheme seems to fit much better purely repulsive systems, as it
incorporates the existence of two singular points on the real axis
\cite{baraml,laifisher,parkfisher}. Yet, a major shortcoming of this
approach was its failure at the critical regime. It is easy to prove
(see below) that tridiagonal $R$ matrices described at the asymptote
by two constant values lead to universal critical exponents
$\sigma=\sigma'=1/2$ at both singularities, which are obviously wrong.
Thus, the above approach fails when one is in close vicinity to
the transition region.

\begin{table*}
\begin{tabular}{|c|c|c|c|}
\hline
 & N4 & N5 & Triangular N2 \\
\hline
$n$ & $nb_n$ & $nb_n$ & $nb_n$\\
\hline
1 & 1                   & 1                     & 1         \\
2 & -21                 & -25                   & -13   \\
3 & 529                 & 757                   & 205   \\
4 & -14457              & -24925                & -3513     \\
5 & 413916              & 860526                & 63116     \\
6 & -12213795               & -30632263                 & -1169197 \\
7 & 368115798               & 1114013874                & 22128177 \\
8 & -11270182473            & -41160109013              & -425493585\\
9 & 349244255338            & 1539411287905             & 8282214430\\
10 & -10926999690716            & -58134505912850           & -162784518218 \\
11 & 344563541226829            & 2212737992414500          & 3224828597398 \\
12 & -10935950490228951         & -84773398978877767            & -64304659129557 \\
13 & 348996298644804045         & 3265709152114882760           & 1289359180917536 \\
14 & -11189659831729226400      & -126396751968240912540        & -25974798852799663 \\
15 & 360221541077745515049      & 4911995555642255534862        & 525411435083794040 \\
16 & -11637415720384495480425       & -191566536035975787182277         & -10665744051246882913 \\
17 & 377133138423022266192030       & 7494404630272576450625728         & 217191426304757630038 \\
18 & -12255532866263525229229458    & -294007038999894901106531809 & \\

\hline
\end{tabular}
\caption{Mayer cluster coefficients $nb_n$ for various models}
\label{nbn}
\end{table*}

A partial solution for this problem was recently found, noticing that for
many of the studied models not only the matrix element approach a
constant but also the asymptotic correction to the constant takes a
universal form, following a $1/n^2$ decay of the elements to their
constant asymptotic value \cite{eli}:
\begin{eqnarray}
B_n  \equiv  R_{n,n}&=&B+b/n^2 \nonumber\\
A_{n+1/2} \equiv  R_{n,n+1}&=& A+a/(n+1/2)^2.\label{eq3}
\end{eqnarray}
Under these circumstances, one is able to analytically calculate the
critical exponents at both fluid termination point (the physical
one, at the ordering transition or at the termination of the super-cooled
fluid, and the nonphysical one on the negative real z-axis).
These exponents depend on the amplitudes of
the $1/n^2$ corrections, and generally deviate from $1/2$. This
approach works satisfactorily for many models and tests well against
the known result for the nonphysical singularity that predicts
universal critical exponents depending on dimensionality alone.
\begin{figure}
\includegraphics[width=7cm,height=5cm,angle=-90]{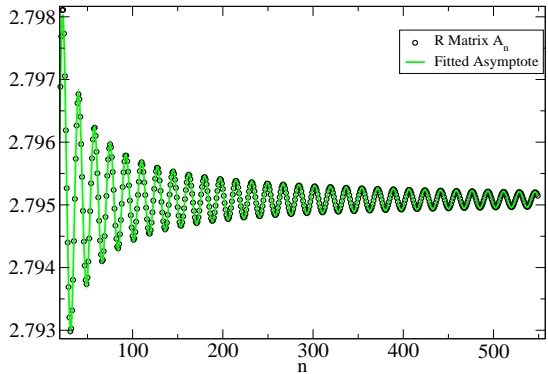}
\includegraphics[width=7cm,height=5cm,angle=-90]{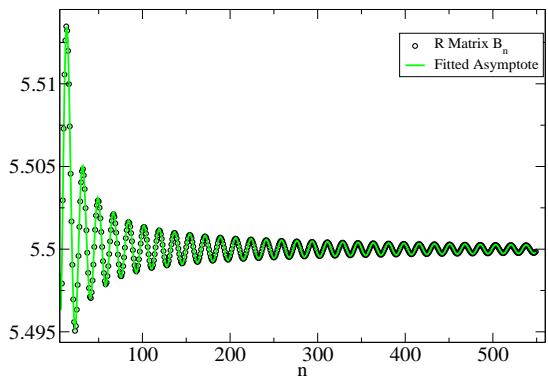}
\caption{Hard hexagons matrix elements, fitted to the form
\eqref{cosform}, with
$A=\sqrt{125}/4$ $a=0.0027$ $a'=-0.063$ $B=5.5$ $b=0.627$ $b'=0.129$
$q=0.36$ } \label{fig-hhl}
\end{figure}
Yet, while many models indeed show this simple $1/n^2$ decay, we have found
out that some other
models exhibit different asymptotic behavior. For example, the $R$
matrix elements of the hard hexagons model
\cite{baxter} are presented in fig~\ref{fig-hhl}. As this is an
exactly solvable
model, one is able to produce a large number of cluster integrals.
Doing so, we note that while the first few elements seem to follow
the $1/n^2$ rule, the asymptotic behavior is quite different. The
matrix elements do converge to two constants as expected, but their
leading asymptotic behavior  follows an oscillatory $1/n$ decay
rather than the above mentioned $1/n^2$.
This finding raises the question of how to deal with
$R$ matrices whose correction deviates from the \eqref{eq3} form.
Moreover, it sheds doubt on the applicability of former results to
other models where only a few cluster integrals are known:
one may argue that the hard hexagons example shows that the $1/n^2$
behavior is only a transient one, and the true asymptotics of all
these models is different. Indeed, extension of the available series
to higher coefficients of the Mayer expansion allowed us to see in a
number of additional models that the seeming $1/n^2$ behavior is
accompanied by additional corrections, including an oscillatory
$\cos(qn)/n$ term that becomes dominant in the asymptote.
We observed such oscillations, for example, for hard-core two-dimensional
square lattice gas with exclusion shell up to second (N2 model),
third (N3 model) and fourth (N4 model)
nearest neighbors.

As this oscillatory term dominates for large $n$, the validity of the
results of \cite{eli} is put in question.
Therefore, we set out to study the effect of this additional correction
term on the critical behavior of the equation of state.
Here we extend the previous result and explore
the case of matrix elements taking the asymptotic form
\begin{eqnarray}
R_{n,n}&=&B+\frac{b}{n^2}+b'\frac{\cos(qn)}{n}\nonumber\\
R_{n,n+1}&=&A+\frac{a}{(n+1/2)^2}+a'\frac{\cos[q(n+1/2)]}{n+1/2}\label{cosform}
\end{eqnarray}
Using an analytical RG-like decimation scheme,
we show that in this case the critical exponents are given by
\begin{eqnarray}\label{expcor}\sigma= \frac{1}{2}\sqrt {1
-\frac{4(2a+b)}{A}- \frac{[2a'\cos(q/2)+b']^2}{[1-\cos(q)]A^2}}\nonumber\\
\sigma '= \frac{1}{2}\sqrt {1 -\frac{4(2a-b)}{A}-
\frac{[2a'\cos(q/2)-b']^2}{[1-\cos(q)]A^2}}
\end{eqnarray}
thus generalizing the results of \cite{eli}. We also discuss the
possible effect of other kinds of corrections, and conclude that
they do not affect the critical exponent as long as the spectrum of
the matrix remains intact. We verify the result by extensive
numerical study of artificial models and by analysis of the exactly
solvable hard-hexagons model. The next-nearest neighbor exclusion
model on a triangular lattice is discussed as an example in which
the spectrum does not remain intact and our approach breaks down.
Finally, we apply our formula to the two models that have been
recently studied by means of Monte-Carlo (MC) simulations
\cite{levin}: the hard-core two-dimensional square lattice gas with
exclusion shell up to fourth (N4 model) and fifth (N5 model) nearest
neighbors.

\section{Analysis}
For the sake of completeness, we start with a brief review of the
approach presented in \cite{eli}. The Mayer cluster integrals provide
a low-$z$ expansion for the density of a fluid:
\begin{equation}
\rho(z)=\sum\limits_{n = 1}^\infty  nb_n z^n,
\end{equation}
where $b_n$ is the $n^{\rm th}$ Mayer cluster integral. It is always
possible (see appendix A for an explicit construction) to define a
tridiagonal symmetric $R$ matrix which satisfies the condition
($n\geq1$)
\begin{equation}\label{eq1} (R^n)_{11}=(-1)^n(n+1)b_{n+1}.
\end{equation}
The density may then be expressed in terms of $R$:
\begin{equation}\label{eq2} \rho(z)=\sum\limits_{n = 1}^\infty  {nb_n z^n
= \sum\limits_{n = 0}^\infty {({ - 1})^n z^{n + 1}( {R^n })_{11}
= z({I + zR})_{11}^{- 1} } }.
\end{equation}
Alternatively, the matrix inversion in the previous equation may be expressed
in terms of the spectrum $\lambda$ of the
$R$ matrix, and the corresponding eigenvectors $\psi(\lambda)$:
\begin{equation}\label{rho_lambda}
\rho(z)=\sum\limits_{\lambda}\frac{\psi_1(\lambda)^2}{z^{-1}+\lambda},
\end{equation}
where $\psi_1(\lambda)$ is the first component of the
$\psi(\lambda)$ vector.

The reciprocals of the eigenvalues of this matrix are the Yang-Lee
zeroes of the grand-canonical partition function. For all purely
repulsive models studied to date, the R matrices are real-valued,
and thus their eigenvalues are also real ($R$ is symmetric by
construction). There is yet no proof that this is indeed the case
for all such models, but construction of $R$ matrices
for dozens of different lattice and continuum purely repulsive
models (see, e.g., \cite{eb06,baraml,eli} and this work) provides
strong evidence for it: in all cases studied the matrix elements
were real to all orders calculated. Furthermore, as mentioned above,
the matrix elements in all models studied adopt a clear asymptotic
pattern, converging quickly to a (real) constant. Therefore the
possibility that some higher order element may become complex seems
improbable.

For these real $R$ matrices the spectrum of the matrix
lies on the real axis in an interval $(-z_t^{-1},z_0^{-1})$ (and the
Yang-Lee zeroes lie on two intervals along the real activity axis:
$z<-z_0$ and $z > z_t$). It follows from \eqref{rho_lambda} that the
density $\rho(z)$ has two singular points at $z$ values for which
$-z^{-1}$ coincides with the spectrum edges of the $R$ matrix,
leading to vanishing of the denominator on the right-hand side. The
critical behavior of the density $\rho(z)$ near the physical and
non-physical singularities is therefore determined by the structure
of the residue $\psi_1(\lambda)$ at the spectrum edges.

For example, we look at a matrix with two constants along the three main
diagonals, $B$ (diagonal) and $A$ (off-diagonal). The eigenvalues are
$\lambda(k)=B+2A\cos(k)\,\,\,\,(0<k<\pi)$ and the eigenvectors
are $\psi_n[\lambda(k)]=\sin(nk)$. The critical points are then
$$-z_0^{-1}=-(B+2A)$$ (corresponding to $k=0$), and
$$z_t^{-1}=2A-B$$
($k=\pi$), where $\psi_1(k)\equiv\psi_1[\lambda(k)]\sim k$ and
$\psi_1(k)\equiv\psi_1[\lambda(k)]\sim (k-\pi)$ respectively.
Expanding the integral in \eqref{rho_lambda} for $z\sim -z_0$ and
$z\sim z_t$ one finds that the density terminates at both ends with
a square-root singularity.

We now consider a general $R$ matrix taking the form
\begin{eqnarray}
B_n\equiv R_{n,n}&=&B+\delta B_n\nonumber\\
A_{n+1/2}\equiv R_{n,n+1}&=&A+\delta A_{n+1/2}.\label{genform}
\end{eqnarray}
The critical behavior is determined by the
long-wavelength, slowly-varying, eigenvectors and therefore
the eigenvalue equation (we treat the nonphysical critical point only,
analysis of physical point is essentially identical)
\begin{equation}\label{SEV}
A_{n-1/2}\psi_{n-1}+B_n\psi_n+A_{n+1/2}\psi_{n+1}=\lambda\psi_n
\end{equation}
may be studied in the continuum limit, taking the form of a
differential equation in the variable $x=kn$.
For the general case \eqref{genform}, the
discrete equations \eqref{SEV} transform into
\begin{equation}\label{diffeq}
f''(x)+f(x)+\frac{[\delta B_n+\delta A_{n-1/2}+\delta
A_{n+1/2}]n^2}{Ax^2}f(x)=0.
\end{equation}
As long as the corrections $\delta B$ and $\delta A$ are small
enough (see below) the spectrum does not change. The eigenvectors,
nevertheless, are modified. In \cite{eli} the $R$ matrix was assumed
to take the form \eqref{eq3}, and then the  differential equation
\eqref{diffeq} is reduced into a Bessel equation. A closed form for
the eigenvectors is available, and one obtains the critical behavior
of the density near the two branch points
$\rho(z_c)-\rho(z)=(z_c-z)^\sigma$ (or $\rho(z)=(z_c-z)^{-\sigma}$
if the density diverges at criticality, such as the case of the
non-physical singularity in $d\leq 2$). The critical exponents are
given by
\begin{equation}\label{exp} \sigma= \frac{1}{2}\sqrt {1 - 4\frac{{2a
- b}}{A}},\,\,\,\,\,\,\,\,\sigma'  = \frac{1}{2}\sqrt {1 - 4\frac{{2a
+ b}}{A}}
\end{equation} where $ \sigma (\sigma') $ is the exponent of the non-physical
(physical) branch point.

This approach, however, cannot be
extended straight-forwardly to study a general correction
to the matrix elements: while for $1/n^2$ corrections \eqref{diffeq} can
be written in terms of $x=kn$ alone, independently of  $k$,
a general correction term results in a $k$-dependent differential equation.
More importantly, considering terms $O(1/n^3)$ in the differential equation
approach leads to an essential singularity at the origin, resulting in
transition layer solutions
and complicated behavior at the origin. These terms indeed show up when one
analyzes real $R$ matrices (see below for the N4 and N5 models).
Third, the mapping to a differential equation relies on the slow variation of
the eigenvectors and is bound to fail for
correction terms of the form \eqref{cosform} that induce an intrinsic
``length''-scale (on the $n$ axis) into the problem.

We thus present here a complementary approach to study the general
correction term, which is based on the idea of renormalization. In
their discrete form, the eigenvalue equations \eqref{SEV} form an
infinite linear system of equations. Since the system is
tridiagonal, it is quite easy to eliminate half of the variables,
e.g. all variables $\psi_n$ for $n$ even. This effectively removes
half of the rows and half of the columns in the matrix, ``tracing
out'' half of the degrees of freedom in the problem. One obtains a
new tridiagonal system of equations, or a renormalized $R$ matrix,
with the same eigenvalues and new vectors $\tilde\psi(k)$ that are
simply related to the former ones $\tilde\psi_n (k)= \psi_{2n-1}(k)$. In
particular, $\tilde\psi_1(k)=\psi_1(k)$. The density
as a function of $z$ is fully determined by the spectrum and
$\psi_1(\lambda)$ through \eqref{rho_lambda}. Thus, the renormalized
$R$ matrix may be utilized to generate the same equation of state
and the same critical behavior as the original one.
Explicitly, the reduced eigenvalue equation after one such decimation
process takes the form (n odd; $A_{n+1/2}=B_n=0$ for $n\leq 0$)
\begin{widetext}\begin{equation}\label{SRG}
\frac{A_{n-3/2}A_{n-1/2}}{\lambda-B_{n-1}}\psi_{n-2}
+(\frac{A_{n-1/2}^2}{\lambda-B_{n-1}}+\frac{A_{n+1/2}^2}{\lambda-B_{n+1}}
+B_n)\psi_n+\frac{A_{n+1/2}A_{n+3/2}}{\lambda-B_{n+1}}\psi_{n+2}=\lambda\psi_n.
\end{equation}\end{widetext}
Accordingly, the $R$ matrix elements transform, under such
decimation, according to
\begin{eqnarray}
B_n'&=&\frac{A_{2n-3/2}^2}{\lambda-B_{2n-2}}+\frac{A_{2n-1/2}^2}{\lambda-B_{2n}}
+B_{2n-1}\\
A_{n+1/2}'&=&\frac{A_{2n-1/2}A_{2n+1/2}}{\lambda-B_{2n}}.
\end{eqnarray}
In the transformed linear system $\tilde\psi_n$ is in fact $\psi_{2n-1}$,
so for a given functional form for $A_n$ and $B_n$
one should change variables $n'\to2n-1$.
Note that the renormalization transformation is $\lambda$-dependent.
Since the density in the vicinity of the critical points is determined by
the spectrum edges only, this poses no difficulty.

As a first demonstration of this RG scheme, one may look at the solvable
case of $1/n^2$ correction. Substituting
$A_n=A+a/n^2$, $B_n=B+b/n^2$ and $\lambda=2A+B$ into \eqref{SRG},
one obtains
\begin{eqnarray}
A_{n+1/2}& \to & A/2+\frac{1}{8n^2}(4a+b)+O(\frac{1}{n^3})\\
B_n & \to & (A+B)+\frac{1}{n^2}(a/2+3b/8)+O(\frac{1}{n^3}).\nonumber
\end{eqnarray}
Clearly, the
spectrum edge, defined by the asymptotic value of
$A_{n-1/2}+A_{n+1/2}+B_n$ to be $-z_0=-(2A+B)^{-1}$ is conserved under
the decimation. Moreover, the correction term $(\delta
A_{n-1/2}+\delta A_{n+1/2}+\delta B_n)/A$ which appears in the
differential equation and determines the critical exponent by
\eqref{exp}, is also stable under the transformation and remains
equal to $(2a+b)/An^2$, as expected.

Applying the same transformation for corrections of the form
$1/n^\alpha$ i.e. $A_n=A+a''/n^{\alpha}$,$B_n=B+b''/n^{\alpha}$, results
in
\begin{equation}\label{powerlaw}
\frac{1}{A}(\delta A_{n-1/2}+\delta A_{n+1/2}+\delta B_n )
\to \frac{1}{2^{\alpha-2}}\frac{2a''+b''}{A}\frac{1}{n^\alpha}.
\end{equation}
Therefore, one may conclude that for $\alpha >2$ the correction term in
the differential equation \eqref{diffeq} is suppressed by successive
applications of the RG decimation transformations.
Therefore, these correction terms are irrelevant in determining the
critical exponents.

We now employ the RG scheme to study the case of main interest:
$1/n$-modulated oscillations, as observed for the hard hexagons model
\begin{eqnarray}
A_{n+1/2}&=&A+a'\cos[q(n+1/2)]/(n+1/2)\nonumber\\
B_n&=&B+b'\cos(qn)/n\label{purecos}.
\end{eqnarray}
The transformation of the differential equation correction term
$(\delta A_{n-1/2}+\delta A_{n+1/2}+\delta B_n)/A$ upon one
decimation step is given by
\begin{widetext}\begin{eqnarray}\label{cosRG}
\frac{1}{A}(\delta A_{n-1/2}+\delta A_{n+1/2}+\delta B_n ) &=&
\frac{2a'\cos(q/2)+b'}{A} \nonumber \\
\to \frac{[2a'\cos(q/2)+b'][1+\cos(q)]\cos(2qn)}{An} &+&
\frac{[2a'\cos(q/2)+b']^2}{2A^2}\frac{1}{n^2}+ O(\cos(2q)/n^2,1/n^3)
\end{eqnarray}\end{widetext}
Obviously, the real-space renormalization process induces a change
in the frequencies $q \to 2q$. In addition, (i) the $\cos(qn)/n$
term is multiplied by a factor of $[1+\cos(q)]$, and three more
terms emerge: (ii) a new $1/n^2$ term, (iii) terms
$O(\cos(2qn)/n^2)$ and (iv) terms $O(1/n^3)$. Iterating this
procedure $N$ times, one obtains from (i) $(2a+b) \to
(2a'+b')\prod\limits_{n = 1}^N [1+\cos(2^{n-1}q)] $. The newly
emerging $1/n^2$ terms (ii) combine to take the form
$\frac{(2a'+b')^2}{16A} \sum\limits_{n = 0}^N \prod\limits_{m = 1}^n
[1+\cos(2^{m-1}q)]^2$. The first term gets exponentially small for
large $N$: $\prod\limits_{n = 1}^N [1+\cos(2^{n-1}q)] \simeq 4^{-N}$
(see Appendix B), and thus could be neglected. The sum over the
products in the second term converges to $[1-\cos(q)]^{-1}$ (see
Appendix B). This second term does affects the critical behavior as it
adds up to the $1/n^2$ terms in the $R$ matrix. The $1/n^3$ terms
(iii) may be neglected as their amplitude decreases: each existing
$1/n^3$ term decreases by factor $2$ upon an RG step, according to
\eqref{powerlaw}. While a new term is being added from the
transformation \eqref{SRG}, the sum of all contributions still
decreases exponentially with the number $N$of RG steps. The
$\cos(qn)/n^2$ terms (iv) transform under decimation in an analogous
way to the original $\cos(qn)/n$ term: they get multiplied by a
factor $1+\cos(q)$ resulting in an exponential decay, and give rise
to new $O(1/n^4)$ terms (analogous to the $O(1/n^2)$ terms generated
from decimation of the $\cos(qn)/n$), as well as faster decreasing
terms. Again, the $1/n^4$ exponentially decrease through decimation
by \eqref{powerlaw} and are therefore neglected. In summary, the net
effect of the $\cos(qn)/n$ term after a large number of RG steps is
the creation of a new $1/n^2$ term. These terms, emerging from the
decimation process, can then be analyzed using the mapping to the
Bessel differential equation as described in \cite{eli}.

Up to this point we treated the pure $\cos(qn)/n$ case. Similar analysis
may be done for the mixed case, where both $\cos(qn)/n$ and $1/n^2$
terms are present (as happens for the physical models to be discussed).
It turns out that the transformation equation \eqref{SRG} does mix the
correction terms, as the numerator of $(A_nA_m)/(\lambda-B_l)$ is
quadratic in the off-diagonal elements. However, the mixed terms
will follow the form $\cos(qn)/n^3$ which can be ignored based on
arguments similar to those presented above for the
$\cos(qn)/n^2$ terms. Multiplicative cross-terms could be relevant
(in RG sense) only if they decay $O(1/n^2)$ or slower. Thus, one may
simply add the original $1/n^2$ terms to those emerging from RG.
Collecting the $O(1/n^2)$ terms originating from both the functional
form of the matrix elements and the decimation process for oscillatory terms,
one obtains the closed form \eqref{expcor} for the critical exponents
in the general case (having both $\cos(qn)/n$ terms and $1/n^2$ terms).

We note that the critical exponents in \eqref{expcor} do not depend
on $B$. This can be readily understood looking at equation
\eqref{rho_lambda}. A change in $B$ results in a constant addition
to the whole spectrum of $R$, without modifying its eigenvectors. Looking
at the expression \eqref{rho_lambda}
for the density, it becomes clear that the effect of a constant added
to the eigenvalues of $R$ on the density is equivalent to
a constant shift in $z^{-1}$.
That is, if the density of the $B=0$ matrix is $\rho_0(z)$,
the density for finite $B$, $\rho(z;B)$, is simply
$$\rho(z; B) = \rho_0(\frac{1}{z^{-1}+B}).$$
Therefore, the value of $B$ affects the location of the critical points,
but do not change the critical exponents.

Finally, we note that if the amplitude of the correction to a
constant matrix is strong enough, one obtains from \eqref{expcor} an
imaginary value for $\sigma$. When this happens, both solutions of
the differential equation \eqref{diffeq} diverge at the origin
\cite{bessel}. Consequently, there are no solutions to the
eigenvalue problem \eqref{SEV} for $k\sim 0$, or $\lambda\sim 2A+B$.
In other words, a perturbation of the constant matrix which is
strong enough to make $\sigma$ imaginary modifies the spectrum of
the matrix, such that the spectrum edge shifts from $2A+B$. In these
cases the critical point is not given by $-z_0=-(2A+B)^{-1}$.
Similarly, whenever $\sigma'$ becomes imaginary, the physical
singularity shifts from $z_t=(2A-B)^{-1}$. In both cases, the
corresponding critical exponents are not given by \eqref{expcor}.
This scenario is realized for the next-nearest neighbor
exclusion model on the triangular lattice (see below).

\section{Numerical Study}

\begin{figure}
\includegraphics[width=8cm,height=6cm,angle=0]{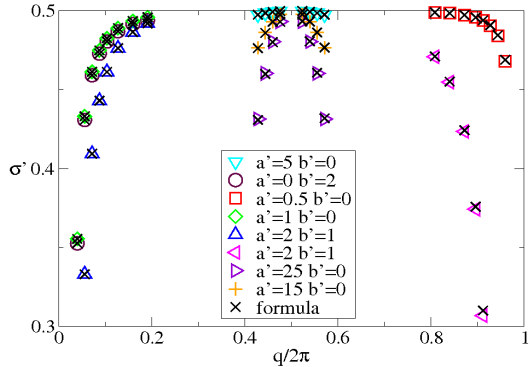}
\caption{Critical exponents as measured from equation of state
\eqref{contfrac}, compared with the exact prediction \eqref{expcor}
for various choices of $\cos(qn)/n$ corrections.}
\label{sig_q}
\end{figure}

In order to test our results, we have constructed various
tridiagonal symmetric $R$ matrices with prescribed matrix-elements
asymptotic form, and compared the prediction \eqref{expcor} with the
critical behavior as measured from the equations of state
calculated by \eqref{eq2} for these models.
First, we looked at matrices obeying
\eqref{purecos}, with the parameters $A=16$ and $B=31$ ($z_t=1$),
and various combinations of $a'$,$b'$ and $q$. We have also modified
$A$ while keeping the other parameters fixed to check the $A$
dependence. For these $R$ matrices, one is able to consider as many
coefficients as desired. Thus, the size of the sub-matrices studied
is considerable, and the matrix inversion of \eqref{eq2} is costly.
Instead, the density $\rho$ can be equivalently calculated using the
continued fraction representation
\begin{widetext}\begin{equation}\label{contfrac} \rho
(z)=\cfrac{1}{R_{1,1}+1/z-R_{1,2}^2\cfrac{1}{R_{2,2}+1/z-R_{2,3}^2
\cfrac{1}{R_{3,3}+1/z-R_{3,4}^2
\cdots
}}}
\end{equation}\end{widetext}
which typically converges rather quickly (except for the immediate
vicinity of the transition point).
Figure ~\ref{sig_q} compares the critical exponent $\sigma'$
obtained by fitting the density as given by \eqref{contfrac} for $z$
close to the termination point $z_t$ with the theoretical prediction
of \eqref{expcor}. The results are in excellent agreement, except
for a few points where the numerical calculation of the density was
difficult due to slow convergence of the continued fraction in the
immediate vicinity of the critical point. We also calculated the
density for $R$ matrices with both $1/n^2$ and $\cos(n)/n$
corrections, i.e. following \eqref{cosform}. The agreement between
the theoretical prediction of \eqref{expcor} and the measured
critical exponent was again excellent. Another special case we
checked was that of a $1/n^3$ correction. This is the most dominant
correction for which we predict no change to the critical exponent
of the const matrix. Using the same constants $A$ and $B$, we looked
at correction amplitudes up to $a''=30$, and verified that the
critical exponent indeed does not change: $\sigma '=0.5$ as
expected.

\begin{table*}
\begin{tabular}{|c|c|c|c|c|c|c|}
\hline
& \multicolumn{2}{|c|}{N4} & \multicolumn{2}{|c|}{N5} & \multicolumn{2}{|c|}{Triangular N2} \\
\hline
$n$ & $B_n$ & $A_n$ & $B_n$ & $A_n$& $B_n$ & $A_n$\\
\hline
1 & 21 & 9.3808315196 & 25 & 11.489125293  & 13 & 6  \\
2 & 17.045454545 & 8.7248457800 & 20.454545455 & 10.570731201 & 10.555555556 & 5.5674871873 \\
3 & 17.024781724 & 8.6098449927 & 20.518434015 & 10.405877593 & 10.550189740 & 5.4798922624  \\
4 & 17.018848337 & 8.5688057487 & 20.535452306 & 10.348827163 & 10.576974307 & 5.4386602176  \\
5 & 17.017061106 & 8.5493273256 & 20.540912637 & 10.322722326 & 10.600981921 & 5.4160504788  \\
6 & 17.016534640 & 8.5385152615 & 20.543142266 & 10.308577639 & 10.615727211 & 5.4037594351  \\
7 & 17.016426427 & 8.5318795598 & 20.544349327 & 10.299999108 & 10.622716309 & 5.3973336220  \\
8 & 17.016464632 & 8.5275092336 & 20.545165481 & 10.294371298 & 10.625196191 & 5.3938592930  \\
9 & 17.016552228 &   & 20.545790085 & & &\\

 \hline
\end{tabular}
\caption{R matrix elements for various models} \label{rmat}
\end{table*}

\section{Applications to physical models}
Analysis of the R matrix as detailed above may be used to predict the
critical behavior of all models with purely repulsive interactions.
Our results apply equally to continuum and lattice models in all dimensions.
Here we demonstrate applications to a number of 2D hard-core lattice
gas models.

For all the models to follow, we have calculated the cluster
integrals to a high order (in order to calculate the $R$ matrix). It
is natural to compare standard series analysis methods
\cite{guttmann} to the results to be obtained from the $R$-matrix.
We have applied the ratio method, Dlog Pad\`{e} and differential
approximants to the models to follow. In general, ratio analysis of
the series provide a rather exact estimate of the non-physical
singularity location $z_0$ and the related $\sigma=1/6$, but says
nothing about the physically relevant $z_t$ and $\sigma'$. Dlog
Pad\`{e} approximants again converge nicely to predict a singularity
at $-z_0$ but show no consistent pole anywhere on the positive real
z-axis. Similar results were obtained using the differential
approximants. Overall, these methods do better then the $R$ matrix
for the nonphysical singularity. The reason for these failures is
the existence of a branch-cut singularity located so close to the
origin, which makes the physical singularity, typically much further
away, undetectable by these methods. The $R$ matrix, which
incorporates the branch-cut naturally, is more successful.

Even though standard series analysis methods are often superior to
the $R$ matrix as a means to analyze the non-physical singularity,
we still include in the following the $R$-matrix results for both
singularities. The reason is that unlike standard methods, $R$
matrix is expected to work equally well for both termination points.
The accuracy of both exponents $\sigma$ and $\sigma'$ depends
roughly equally on the quality of the fitting parameters describing
the asymptotic behavior of the matrix elements. Thus, our $R$ matrix
results for $\sigma$ should not be taken as the yardstick for
measuring $R$ matrix vs. Dlog Pad\`{e}, but rather as a measure of
the accuracy of the $R$ matrix itself, as one expects the same
degree of accuracy for both exponents calculated.

\subsection{Hard hexagons model}
The hard hexagons model (lattice gas on on a triangular lattice with
nearest-neighbors exclusion) was solved exactly by Baxter
\cite{baxter}. This allows us to calculate many cluster coefficients
and matrix elements. The density in this model is given exactly by
the relation \cite{joyce}
\begin{widetext}\begin{eqnarray}\label{hhl_nbn}
&&\rho^{11}(\rho-1)z^4-\rho^5(22\rho^7-77\rho^6+165\rho^5-220\rho^4+165\rho^3-66\rho^2+13\rho -1)z^3  \nonumber \\
&+&\rho^2(\rho-1)^2(119\rho^8-476\rho^7+689\rho^6-401\rho^5-6\rho^4+125\rho^3-63\rho^2+13\rho -1)z^2  \nonumber \\
&+&(\rho
-1)^5(22\rho^7-77\rho^6+165\rho^5-220\rho^4+165\rho^3-66\rho^2+13\rho
-1)z+\rho (\rho -1)^{11}=0.
\end{eqnarray}\end{widetext}
Using this relation, one is able to expand the density in power
series of the activity $z$ and extract the cluster integrals $nb_n$.
Employing infinite-precision integer computation we extended the 24
elements calculated in \cite{joyce} to 1100 elements, enabling the
construction of the first 550 diagonal and off-diagonal elements of
the R matrix \cite{website}. These allowed unambiguous determination
of the asymptotic form of these elements. One can observe in
figure~\ref{fig-hhl} clear oscillations of the matrix elements.
Therefore application of the formula presented in
\cite{eli}, which is based on a $O(n^{-2})$ correction term, was
doubtful. Based on the analysis above and the extended formula
\eqref{expcor}, one may calculate the critical exponent from fitting
the matrix elements of the hard hexagons model. This results in
$\sigma '=0.6662$ where the exact result is $\sigma'=2/3$. Note that
the early version of \eqref{expcor} as presented in \cite{eli}
gives in this case $\sigma' =0.6902$. The result for the
nonphysical critical exponent calculated based on our $R$ matrix analysis and
\eqref{expcor} is $\sigma =0.1655$, which compares well to
the exact universal result $\sigma=1/6$ \cite{baraml,laifisher,parkfisher}.

\subsection{Triangular lattice N2 model}
Next, we study the triangular lattice N2 model (exclusion up to
the next-nearest-neighbor). This model was long ago investigated,
and early studies suggested that the phase transition is first
order \cite{orban,runnels,nisbet}. However, later transfer matrix
analysis \cite{bartelt}, and recent exhaustive MC results \cite{zhang}
concluded that the model undergoes a second order
phase transition at $\mu_c=1.75682(2)$ and critical density
$\rho_c=0.180(4)$, and is believed to be part of the $q=4$ Potts
universality class, with $\sigma'=1/3$.

\begin{figure}
\includegraphics[width=7cm,height=5cm,angle=-90]{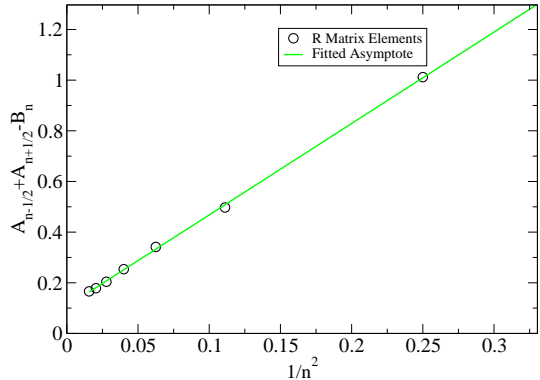}
\caption{Triangular N2 Matrix elements. The difference
$A_{n+1/2}-B_n+A_{n+1/2}$ extrapolates to 0.107(1), much lower than
$1/z_t=0.1726$.
The slope with respect to $n^{-2}$ is 3.61(1), much larger than $A/4$.
These two observations are consistent with a spectrum edge shift.
} \label{fig-trin2c}
\end{figure}

We used the transfer matrix method to obtain an exact expansion of
the partition function in powers of the activity. We have
constructed transfer matrices for strips with width up to $M=26$
(number of symmetry reduced states in the $M=26$ matrix is 730100).
We then constructed the exact low-$z$ power series expansion for the
density $\rho(z)$, the first 17 coefficients of which are identical
with their bulk values (the cluster integrals for the models
considered henceforth and the resulting $R$ matrices are given in
tables~\ref{nbn},~\ref{rmat}). The difference
$A_{n-1/2}+A_{n+1/2}-B_n$ should converge to $2A-B=z_t^{-1}$. In the
absence of oscillatory terms, the slope of this difference against
$1/n^2$ determines the critical exponent by \eqref{expcor}. As seen
in figure~\ref{fig-trin2c} the matrix elements are well fitted, with
$2A-B=0.107(1)$ and $2a-b=3.61(1)$, and extrapolation of $A_n$ alone
gives $A=5.382$. Therefore, in this case analysis of the R matrix
shows clearly that $4(2a-b)/A\simeq 2.7 >1$ which means that
\eqref{exp} will lead to an imaginary $\sigma'$. As discussed above,
in such cases the above analysis breaks down as the spectrum edge
shifts from $2A\pm B$. Indeed, for this model the critical activity
as determined by MC studies, $z_t =5.794$  \cite{zhang}, deviates
significantly from $(2A-B)^{-1} = 9.35$, clearly demonstrating the
spectrum edge shift.

\subsection{Square lattice N4 model}
Having tested the limits of the method, we move on to apply it and
examine models in which the critical behavior is not known. The N4
model on a square lattice (hard-core exclusion of all neighbors up
to the $4^{th}$ order) was first studied using transfer matrix
methods \cite{orbanphd,nisbet}. Recently, it was revisited,
employing MC simulations \cite{levin}. It is believed to
undergo a second order fluid-solid transition of the Ising
universality class. The critical chemical potential was found to be
$\mu _c=4.705$ with a critical density $\rho_c=0.110$ \cite{levin},
where the closest packing density is $\rho_{cp}=0.125$.

\begin{figure}
\includegraphics[width=7cm,height=5cm,angle=-90]{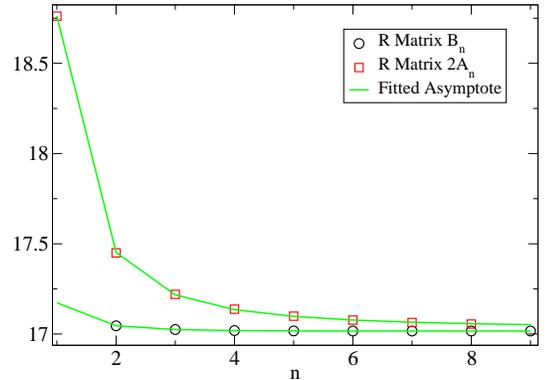}
\caption{N4 Matrix elements. The diagonal terms are fitted to
the functional form \eqref{cosform}:
$B=17.0121$, $b=0.19$, $b'=0.029$,
$q=0.295$. The off diagonal term fit well \eqref{eq3} with an
added cubic correction $a''/n^3$:
$2A=17.0316$, $2a=1.634$, $2a'=0$, $2a''=0.0957$. }
\label{fig-n4}
\end{figure}

Here too, we used the transfer matrix method to obtain an exact
expansion of the partition function and expand the density in powers
of the activity. We have constructed transfer matrices for strips
with width up to $M=37$. Employing translational and inversion
symmetries, the number of symmetry reduced states in the $M=37$
matrix is 4137859. Using this matrix, we obtained the first 18
coefficients that are identical with the bulk values. The
diagonal matrix elements take the form $B+b/n^2+b'\cos(qn+\phi)/n$,
while the off-diagonal ones exhibit no visible oscillations, and are
well fitted by the cubic form $A+a/n^2+a''/n^3$ (see
figure~\ref{fig-n4}). Based on the fit parameters, one is able to
predict the non-physical singularity location $-z_0=-0.0294$, which
compares well with the value we obtained from direct ratio analysis
of the series $-z_0=-0.029374(1)$.
The critical exponent at this singularity is
calculated by \ref{expcor} to be $\sigma = 0.1891$, close to
the exact universal value $\sigma=1/6$.

Looking at the physical singularity, one observes $2A-B=0.015(5)$,
i.e., $\mu_c=4.2(4)$, barely consistent with the result of
\cite{levin}. While the accuracy in determining the critical
activity is low, the critical density can be determined to much
better accuracy
$\rho_c=0.112(1)$, in good agreement with the MC results. It is
remarkable that we are able to determine to such accuracy the
critical density at the fluid-solid transition based on the
low-density behavior of the fluid alone. The critical exponent may
be found by \eqref{expcor} to be $\sigma'=0.28(6)$.
Thus, based on our analysis of the cluster integrals
we can quite safely exclude the possibility of the Ising
universality  class, where $\sigma'=1$. The latter result contradicts the
numerical observations of \cite{levin}. Detailed numerical
studies of this model aimed at an accurate calculation of the
critical exponents are required to settle this discrepancy.

\subsection{Square lattice N5 model}
\begin{figure}
\includegraphics[width=7cm,height=5cm,angle=-90]{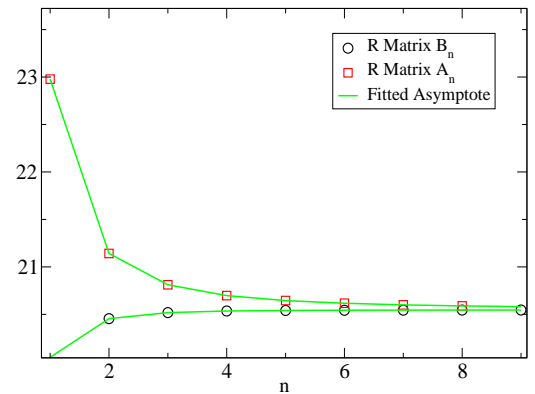}
\caption{N5 Matrix elements. Diagonal and off-diagonal terms fit
well \eqref{eq3} with an added cubic correction $b''/n^3$
($a''/n^3$) : $B=20.547$, $b=-0.0166$, $b''=-0.706$. $2A=20.553$,
$2a=2.28$, $2a''=0.143$. } \label{fig-n5}
\end{figure}

Finally, we look at  the N5 model on a square lattice (hard-core
exclusion of all neighbors up to the $5^{th}$ order). This model
was also recently studied using MC simulations \cite{levin}
and found to undergo a weak first order transition at
$\mu_c=5.554$. Again, we calculated 18 cluster coefficients using the
transfer matrix method up to $M=37$. In this case, one observes no
oscillations, but the $R$ matrix elements exhibit a strong
third-order correction term: $A_n=A+a/n^2+a''/n^3$ and
$B_n=B+b/n^2+b''/n^3$ (see figure~\ref{fig-n5}). While the third
order term is stronger than the second-order one in the regime
studied, our RG analysis allows us to conclude that the $1/n^3$
correction does not change the critical exponents and we can use
\eqref{exp}. The non-physical exponent $\sigma$ calculated from the
above parameters, $\sigma=0.1718$ is in reasonable agreement with
the exact universal result $1/6$. Similar calculation for the
physical singularity yields $\sigma=0.1621$. The accuracy of the
latter result  might suffer from the lack of insufficient cluster
integrals. However, one can safely say that the diagonal $1/n^2$
amplitude $b$ is small, and thus the physical exponent $\sigma '$
would not deviate much from $\sigma$, and should satisfy
$\sigma'\simeq 1/6$. The critical activity $z_t=(2A-B)^{-1}$ is
estimated to be $z_t=166$, but is highly
sensitive to small errors in $A$ and $B$ and might be very well
equal or higher than the one
reported in \cite{levin} ($z_c=258$). If $z_t>z_c$ then the critical point we
found corresponds to the termination of the super-cooled fluid
phase. This scenario is discussed in \cite{eliasher} and was
suggested to be related to a glass transition \cite{eliasher,eli}.

\section{Conclusion}

The $R$ matrix representation of the Mayer cluster integrals
converges very quickly to its asymptotic form. It therefore provides
a powerful tool for extrapolating the low-$z$ expansion of the fluid
equation of state to cover the full fluid regime. In this work we
analyze the analytic properties of this equation of state in the
vicinity of the critical points. It is shown that not only the
location of the critical points, but also the critical exponents can
be determined if one identifies correctly the asymptotic behavior of
the $R$ matrix elements. A number of correction forms are analyzed,
most of which are shown by RG arguments to be irrelevant for the
critical behavior. Thus, we provide an exact formula for the
critical exponents, depending on a relatively few parameters
characterizing the functional dependence of the matrix elements.
Application of this method to a number of lattice-gas models results
in partial agreement with recent MC studies. Analysis of the
discrepancies through an extensive MC study is left for future work.

\begin{acknowledgements}
We thank Asher Baram for numerous discussions and for critical
reading of the manuscript.
\end{acknowledgements}

\section{Appendix A: Construction of the $R$-matrix}

Here we give an explicit recursive construction of a tridiagonal symmetric
$R$ matrix that satisfies \eqref{eq1}.
First, assign
$$
R_{11}=-2b_2.
$$
Assuming all elements $R_{ij}$ are known for $1\leq i,j \leq m$
(and \eqref{eq1} is satisfied for $n\leq 2m$), we construct $R_{m,m+1}$ and
$R_{m+1,m+1}$ as follows:

Define $P$ to be the $m\times m$ leading submatrix of $R$, i.e., the first
$m$ rows and first $m$ columns of $R$. The next off-diagonal element is
given by
$$
R_{m,m+1}^2 = R_{m+1,m}^2 =
\frac{(2m+1)b_{2m+1} - (P^{2m})_{11}}{(P^{m-1})_{1m}^2}.
$$
Now define $Q$ to be the $(m+1)\times(m+1)$ leading submatrix of
$R$, with zero as its $m+1,m+1$ element. The next diagonal element is
then given by
$$
R_{m+1,m+1} = \frac{-(2m+2)b_{2m+2} - (Q^{2m+1})_{11}}{(Q^{m})_{1,m+1}^2}.
$$

It is easy to see by explicit multiplication that the submatrix up to row
and column $m$ satisfy \eqref{eq1} up to $n=2m$. Further matrix elements
do not affect $(R^j)_{11}$ for $j\leq n$. Therefore, each additional
cluster integral allows for one additional $R$ matrix elements. It should be
pointed out that the above process is exponentially sensitive to errors.
This means that if one is interested in matrices with $m>5$ or so,
the cluster integrals used should be exact or at least known to high accuracy.
In addition, the actual construction of $R$ matrices should
generally be done using high-accuracy arithmetics to avoid build-up of
round-off errors.

\section{Appendix B}
We first show that
\begin{equation}\label{id1}
\prod\limits_{j=1}^N[1+\cos(2^{j-1}q)]^2 \simeq 4^{-N},\quad\quad
(N\to\infty).
\end{equation}
Taking the logarithm of the product, one obtains
$2\sum\limits_{j=1}^N \ln[1+\cos(2^{j-1}q)]$. It is easy to see that
for $q/(2\pi)$ irrational, the sequence $2^j q (\mod 2\pi)$ is
uniformly dense in $(0,2\pi)$. Thus, in the limit $N\to\infty$ the
sum may be replaced by an integral
\begin{equation}
2\sum\limits_{j=1}^N \ln[1+\cos(2^{j-1}q)]= 2N\int\limits_{0}^{2\pi}
\ln[1+\cos(x)]dx=-2N\ln(2).\nonumber
\end{equation}
Exponentiating the result, one reveals \eqref{id1}.

Secondly, we show that
\begin{equation}\label{proof1}
f(q)=1+\sum\limits_{i=1}^\infty\prod\limits_{j=1}^i[1+\cos(2^{j-1}q)]^2
=\frac{2}{1-\cos(q)}.
\end{equation}
It follows from the definition that $f(q)$ satisfies
$f(q)-1=[1+\cos(q)]^2f(2q)$. This recursion rule is indeed satisfied
by
 $f(q)=2/[1-\cos(q)]$.
All left to be shown is that there is no other (continuous) solution. Assume
there exist two different solutions $f_1(q)$ and $f_2(q)$. Their difference
$\delta f(q)=f_1(q)-f_2(q)$ then satisfies
\begin{equation}\label{iter}
\delta f(q)=[1+\cos(q)]^2\delta f(2q) =\delta f(2^n
q)\prod\limits_{j=1}^n[1+\cos(2^{j-1}q)]^2
\end{equation}

Let $q/(2\pi)$ be irrational.
$\delta f$ is continuous, thus for each $\epsilon$ there exists
$\delta$ such that $|q_1-q|<\delta \to |f(q_1)-f(q)|<\epsilon$.
Again we use the fact that the sequence $2^j q (\mod 2\pi)$ is
uniformly dense in $(0,2\pi)$ to deduce that there exists also
$N$ such that $|2^N q (\mod 2\pi)- q|<\delta$ and thus
$|\delta f(2^N q)- \delta f(q)|<\epsilon$. In fact there are infinitely
many such $N$'s,
so one may find $N$ as large as required to satisfy the latter inequality,
while at the same time satisfying \eqref{id1}.
Employing \eqref{iter} one finds
\begin{equation}
\delta f(q)=\delta f(2^N q)\prod\limits_{j=1}^N[1+\cos(2^{j-1}q)]^2
\sim 4^{-N} \delta f(2^N q).
\end{equation}
That is, $|f(2^N q)-f(q)| \simeq (4^N-1)|f(q)| >>\epsilon$ in
contradiction to the abode,
unless $\delta f (q)=0$. Since this is true for all irrational
$q/(2\pi)$, the function must
vanish identically if continuous. Q.E.D.


\begin{thebibliography}{50}
\bibitem{onsager}
L. Onsager, Ann. N.Y. Acad. Sci. \textbf{51}, 627 (1949).

\bibitem{zwanzig}
R. Zwanzig, J. Chem. Phys. \textbf{39}, 1714 (1963).

\bibitem{alderdisks}
B.J. Alder and T.E. Wainwright, Phys. Rev. \textbf{127},  357 (1962).

\bibitem{wood}
W.W. Wood and J.D. Jacobson,  J. Chem. Phys. \textbf{27}, 1207 (1957).

\bibitem{alder}
B.J. Alder and T.E Wainwright, J. Chem. Phys. \textbf{33}, 1439 (1960).

\bibitem{hoover}
W.G. Hoover and F.H. Ree,  J. Chem. Phys. \textbf{49}, 3609 (1968).

\bibitem{michels}
P.J. Michels and N.J. Trappaniers, Phys. Lett. A \textbf{104}, 425 (1984).

\bibitem{hexatic1}
D.R. Nelson and B.I. Halperin, Phys. Rev. \textbf{B 19}, 2457 (1979).

\bibitem{hexatic2}
A.P. Young, Phys. Rev. \textbf{B 19}, 1855 (1979).

\bibitem{Groeneveld}
J. Groeneveld,  Phys. Lett. \textbf{3}, 50 (1962).

\bibitem{baramr}
A. Baram and J.S. Rowlinson,  J. Phys. A {\bf23}, L399  (1990).

\bibitem{baramf}
A. Baram and M.J. Fixman, J. Chem. Phys. {\bf101}, 3172 (1994).

\bibitem{baramr2}
A. Baram and J.S. Rowlinson, Mol. Phys. {\bf74}, 707 (1991).

\bibitem{eb06}
E. Eisenberg and A. Baram, Phys. Rev. \textbf{E 73}, 025104(R) (2006).

\bibitem{baraml}
A. Baram and M. Luban, Phys. Rev. {\bf A 36}, 760 (1987).

\bibitem{laifisher}
S.N. Lai and M.E. Fisher, J. Chem. Phys. {\bf103}, 8144  (1995).

\bibitem{parkfisher}
Y. Park and M.E. Fisher, Phys. Rev. \textbf{E 60}, 6323  (1999).

\bibitem{eli}
E. Eisenberg and A. Baram, PNAS {\bf 104}, 5755 (2007).

\bibitem{baxter}
R.J. Baxter, J. Phys. A: Math. Gen. {\bf13}, L61 (1980).

\bibitem{levin}
H.C.M. Fernandes, J.J. Arenzon and  Y. Levin, J. Chem. Phys.
\textbf{126}, 114508 (2007).

\bibitem{bessel}
T.M. Dunster, SIAM J. Math. Anal. \textbf{21}, 995-1018 (1990).

\bibitem{guttmann}A.J. Guttmann, {\it Phase Transition and Critical Phenomena},
vol. 13, ed. C. Domb and J. Lebowitz (New York: Academic)

\bibitem{joyce}
G.S. Joyce, Phil. Trans. Royal Soc. London \textbf{A 325}, 643-702
(1988).

\bibitem{website}
The exact 1100 cluster integrals $nb_n$, and the first 550 $R$
matrix elements are available on-line at
http://star.tau.ac.il/$\sim$eli/Rmat .

\bibitem{orban}
J. Orban and A. Bellemans, J. Chem. Phys. \textbf{49}, 363 (1968).

\bibitem {runnels}
L.K. Runnels, J.R. Craig, and H.R. Stereiffer, J. Chem. Phys.
\textbf{54}, 2004 (1971).

\bibitem{nisbet}
R.M. Nisbet and I.E. Farquhar, Physica (Amsterdam) \textbf{76}, 283
(1974).

\bibitem{bartelt}
N.C. Bartelt and T.L. Einstein, Phys. Rev. \textbf{B 30}, 5339 (1984).

\bibitem{zhang}
W. Zhang and Y. Deng, Phys Rev \textbf{E 78}, 031103 (2008).

\bibitem{orbanphd}
J. Orban, Ph.D. thesis, Universite Libre de Bruxelles (1969).

\bibitem{eliasher}
E. Eisenberg and A. Baram, Europhys. Lett. \textbf{71}, 900 (2005)

\end{thebibliography}
\end{document}